\documentclass[12pt]{article}
\usepackage[table]{xcolor}
\usepackage{colortbl}
\definecolor{lightgray}{gray}{0.9}
\usepackage{graphicx}
\usepackage{lscape}
\usepackage{citesort}
\usepackage{amssymb}
\usepackage{appendix}
\usepackage{multirow}

\begin{document}
\def\qq{\langle \bar q q \rangle}
\def\uu{\langle \bar u u \rangle}
\def\dd{\langle \bar d d \rangle}
\def\sp{\langle \bar s s \rangle}
\def\GG{\langle g_s^2 G^2 \rangle}
\def\Tr{\mbox{Tr}}
\def\figt#1#2#3{
        \begin{figure}
        $\left. \right.$
        \vspace*{-2cm}
        \begin{center}
        \includegraphics[width=10cm]{#1}
        \end{center}
        \vspace*{-0.2cm}
        \caption{#3}
        \label{#2}
        \end{figure}
    }

\def\figb#1#2#3{
        \begin{figure}
        $\left. \right.$
        \vspace*{-1cm}
        \begin{center}
        \includegraphics[width=10cm]{#1}
        \end{center}
        \vspace*{-0.2cm}
        \caption{#3}
        \label{#2}
        \end{figure}
                }

\def\ds{\displaystyle}
\def\beq{\begin{equation}}
\def\eeq{\end{equation}}
\def\bea{\begin{eqnarray}}
\def\eea{\end{eqnarray}}
\def\beeq{\begin{eqnarray}}
\def\eeeq{\end{eqnarray}}
\def\ve{\vert}
\def\vel{\left|}
\def\ver{\right|}
\def\nnb{\nonumber}
\def\ga{\left(}
\def\dr{\right)}
\def\aga{\left\{}
\def\adr{\right\}}
\def\lla{\left<}
\def\rra{\right>}
\def\rar{\rightarrow}
\def\lrar{\leftrightarrow}
\def\nnb{\nonumber}
\def\la{\langle}
\def\ra{\rangle}
\def\ba{\begin{array}}
\def\ea{\end{array}}
\def\tr{\mbox{Tr}}
\def\ssp{{\Sigma^{*+}}}
\def\sso{{\Sigma^{*0}}}
\def\ssm{{\Sigma^{*-}}}
\def\xis0{{\Xi^{*0}}}
\def\xism{{\Xi^{*-}}}
\def\qs{\la \bar s s \ra}
\def\qu{\la \bar u u \ra}
\def\qd{\la \bar d d \ra}
\def\qq{\la \bar q q \ra}
\def\gGgG{\la g^2 G^2 \ra}
\def\q{\gamma_5 \not\!q}
\def\x{\gamma_5 \not\!x}
\def\g5{\gamma_5}
\def\sb{S_Q^{cf}}
\def\sd{S_d^{be}}
\def\su{S_u^{ad}}
\def\sbp{{S}_Q^{'cf}}
\def\sdp{{S}_d^{'be}}
\def\sup{{S}_u^{'ad}}
\def\ssp{{S}_s^{'??}}

\def\sig{\sigma_{\mu \nu} \gamma_5 p^\mu q^\nu}
\def\fo{f_0(\frac{s_0}{M^2})}
\def\ffi{f_1(\frac{s_0}{M^2})}
\def\fii{f_2(\frac{s_0}{M^2})}
\def\O{{\cal O}}
\def\sl{{\Sigma^0 \Lambda}}
\def\es{\!\!\! &=& \!\!\!}
\def\ap{\!\!\! &\approx& \!\!\!}
\def\md{\!\!\!\! &\mid& \!\!\!\!}
\def\ar{&+& \!\!\!}
\def\ek{&-& \!\!\!}
\def\kek{\!\!\!&-& \!\!\!}
\def\cp{&\times& \!\!\!}
\def\se{\!\!\! &\simeq& \!\!\!}
\def\eqv{&\equiv& \!\!\!}
\def\kpm{&\pm& \!\!\!}
\def\kmp{&\mp& \!\!\!}
\def\mcdot{\!\cdot\!}
\def\erar{&\rightarrow&}
\def\olra{\stackrel{\leftrightarrow}}
\def\ola{\stackrel{\leftarrow}}
\def\ora{\stackrel{\rightarrow}}

\def\simlt{\stackrel{<}{{}_\sim}}
\def\simgt{\stackrel{>}{{}_\sim}}


\title{
         {\Large
                 {\bf
                      Investigation  of the semileptonic transition of the $B$   into the orbitally excited  charmed tensor meson 
                 }
         }
      }

\author{\vspace{1cm}\\
{\small K. Azizi$^1$ \thanks
{e-mail: kazizi@dogus.edu.tr }\,\,,\,\,H. Sundu$^2$ \thanks {e-mail: hayriye.sundu@kocaeli.edu.tr}\,\,,\,\,S. \c{S}ahin$^2$ \thanks {e-mail: 095131004@kocaeli.edu.tr}}  \\
{\small $^1$ Department of Physics, Do\u gu\c s University,
Ac{\i}badem-Kad{\i}k\"oy, 34722 Istanbul, Turkey }\\
{\small $^2$ Department of Physics , Kocaeli University, 41380
Izmit, Turkey} }
\date{}

\begin{titlepage}
\maketitle
\thispagestyle{empty}

\begin{abstract}
The transition form factors of the semileptonic $B \rightarrow
D_2^*(2460)\ell\overline{\nu}\,\,\,\,(\ell=\tau,\mu,e)$ decay
channel are calculated within the framework of the three-point QCD
sum rules. The fit functions of the form factors  are then used to
estimate the total decay width and branching ratio of this
transition.  The order of branching ratio shows that this channel
can be detected at LHCb. 
\end{abstract}

~~~PACS number(s): 11.55.Hx, 13.20.He, 14.40.Lb
\end{titlepage}

\section{Introduction}

As it is well known, the semileptonic decays of $B$ meson are  very promising tools in constraining the standard model  parameters,  determination of  the elements of the Cabibbo-Kobayashi-Maskawa (CKM) matrix, 
 understanding  the origin of the CP violation and  looking  for new physics  effects. Over the last few years,  the radially excited charmed mesons have been in the focus of much attention both theoretically and experimentally.
 In 2010, BaBar Collaboration reported their  isolation of  a number of orbitally excited  charmed mesons \cite{bab}. This report has stimulated the theoretical works devoted to the semileptonic decays of 
 $B$ meson into the orbitally excited charmed meson (for instance see \cite{damir,segovia,fazio,colang} and references therein). As the decays of $B$ meson  into orbitally excited
charmed mesons can provide a substantial contribution to the total semileptonic decay width, such processes deserve more detailed studies.  Moreover, a  better knowledge on  these transitions  can help us
  in the analysis of signals and backgrounds
of inclusive and exclusive decays of $b$-hadrons.

In this article, we calculate the transition form factors of the semileptonic decays of $B \rightarrow
D_2^*(2460)\ell\overline{\nu}$ in the framework of the three-point QCD sum rules. This approach is one of the attractive and applicable nonperturbative tools
 to hadron physics based on the QCD Lagrangian \cite{shifman}. 
As the $D_2^*(2460)$ is a tensor meson containing derivatives in its  interpolating current, we start our calculations in the
 coordinate space then we apply the Fourier transformation to go to the momentum space. Based on the general philosophy of the method, to suppress the contributions of the higher states and continuum, we finally 
apply the Borel transformation and continuum subtraction which bring some auxiliary parameters whose working regions are determined demanding some criteria. The transition form factors are then used to calculate
the decay width and branching ratio of the semileptonic decay channel under consideration. 

The BaBar Collaboration has recently measured the  ratios for the branching fractions of the $B$ to charmed pseudoscalar $D$ and vector $D^*$ 
mesons at $\tau$ channel to those of
 the $e$ and $\mu$ channels \cite{Lees}.
The obtained results deviate at the  level of 3.4$~\sigma$ from the  existing theoretical predictions in SM   \cite{Lees,Fajfer}. 
Hence, there is a possibility  
that the  semileptonic transitions  containing  heavy $b$ and $c$ quarks and the $\tau$ 
lepton are bring out the effects of particles with large couplings to the heavier fermions \cite{colangg}. Determination of these ratios of the branching fractions 
in $B$ to charmed tensor $D_2^*$
channel can also be important from this point of view whether these anomalous in the pseudoscalar and vector channels exist in the tensor channel or not. We will be able to answer this question when 
having the experimental data in this channel. By the aforementioned experimental progress on the identification and spectroscopy of the orbitally excited charmed mesons as well as the  developments at LHC
and by considering the orders of the branching ratios in the tensor channel, we hope it will be possible  in near future.

This article is arranged as follows. We derive the QCD sum rules
for the form factors defining the semileptonic $ B \rightarrow D_2^*(2460)\ell\overline{\nu}
$ transition in section 2. The last section is devoted to the numerical analysis of the form factors,  calculations of the branching ratios of the transition  under consideration at different
lepton channels  as well as our concluding remarks.

\section{QCD sum rules for transition form factors of $ B \rightarrow D_2^*(2460)\ell\overline{\nu }$}
This section is dedicated to  calculation of the form factors of
the $B \rightarrow D_2^*(2460)\ell\overline{\nu}$ transition
applying the QCD sum rules technique. The starting point is to
consider the following tree-point correlation function:
\begin{eqnarray}\label{correl.func.101}
\Pi _{\mu\alpha\beta}(q^2)=i^2\int d^{4}x\int d^{4}y
e^{-ip.x}e^{ip'.y}{\langle}0\mid {\cal
T}\Big[J_{\alpha\beta}^{D_2^*}(y) J^{tr}_{\mu}(0)
J^{B^{\dag}}(x)\Big]\mid0{\rangle},
\end{eqnarray}
where, ${\cal T}$ is the time ordering operator and
$J^{tr}_{\mu}(0)=\overline{c}(0)\gamma_{\mu}(1-\gamma_5)b(0)$ is
the transition current. Also, the interpolating current of the $B$
and $D_2^*(2460)$  mesons are written in terms of the quark fields
as
\begin{eqnarray}\label{pseudscalarcurrent}
 J^{B}(x)=\overline{u}(x)\gamma_5b(x)
\end{eqnarray}
\begin{eqnarray}\label{tensorcurrent}
J_{\alpha\beta}^{D_2^*}(y)=\frac{i}{2}\left[\bar u(y)
\gamma_{\alpha} \olra{\cal D}_{\beta}(y) c(y)+\bar u(y)
\gamma_{\beta} \olra{\cal D}_{\alpha}(y) c(y)\right],
\end{eqnarray}
where the $ \olra{\cal D}_{\beta}(y)$ denotes the four-derivative
with respect to $y$ acting on the left and right, simultaneously
and is given as
\begin{eqnarray}\label{derivative}
\olra{\cal D}_{\beta}(y)=\frac{1}{2}\left[\ora{\cal D}_{\beta}(y)-
\ola{\cal D}_{\beta}(y)\right],
\end{eqnarray}
with,
\begin{eqnarray}\label{derivative2}
\overrightarrow{{\cal
D}}_{\beta}(y)=\overrightarrow{\partial}_{\beta}(y)-i
\frac{g}{2}\lambda^aA^a_\beta(y),\nonumber\\
\overleftarrow{{\cal
D}}_{\beta}(y)=\overleftarrow{\partial}_{\beta}(y)+
i\frac{g}{2}\lambda^aA^a_\beta(y),
\end{eqnarray}
where,  $\lambda^a$ are the Gell-Mann matrices and $A^a_\beta(x)$
is the external  gluon fields. These fields are expressed in terms
of the gluon field strength tensor, using the Fock-Schwinger gauge
($x^\beta A^a_\beta(y)=0$),
\begin{eqnarray}\label{gluonfield}
A^{a}_{\beta}(y)=\int_{0}^{1}d\alpha \alpha y_{\nu}
G_{\nu\beta}^{a}(\alpha y)= \frac{1}{2}y_{\nu}
G_{\nu\beta}^{a}(0)+\frac{1}{3}y_\eta y_\nu {\cal D}_\eta
G_{\nu\beta}^{a}(0)+...
\end{eqnarray}

Following the general idea of the QCD sum rule approach, the
aforementioned correlation function is calculated via two
different ways: once in terms of hadronic degrees of freedom
called phenomenological or physical side  and, the second, in
terms of QCD degrees of freedom called theoretical or QCD side.
 By matching  these two representations, the QCD sum rules
 for the form factors are obtained. To
 stamp down the contributions of the higher states and continuum,
 we will apply double Borel transformation with respect to the
 momentum squared of the initial and final states and will use the quark-hadron duality assumption.

\subsection{The phenomenological side}

On the  phenomenological side, the correlation function is
obtained inserting two complete sets of intermediate states with
the same quantum numbers as the interpolating currents $J^B$ and
$J^{D_2^*}$ into Eq. (\ref{correl.func.101}). After performing
four-integrals over $x$ and $y$, we get
\begin{eqnarray}\label{phen1}
\Pi _{\mu\alpha\beta}^{phen}(q^2)=\frac{{\langle}0\mid  J
_{\alpha\beta}^{D_2^*}(0) \mid
D_2^*(p',\epsilon)\rangle{\langle}D_2^*(p',\epsilon)\mid J
_{\mu}^{tr}(0) \mid B(p)\rangle \langle B(p)\mid
J_{B}^{\dag}(0)\mid
 0\rangle}{(p^2-m_B^2)(p'^2-m_{D_2^*(2460)}^2)}
&+& \cdots,\nonumber\\
\end{eqnarray}
 where $\cdots$ represents  contributions of  the higher states and continuum, and
 $\epsilon$ is the polarization tensor of the $D_2^*(2460)$ tensor meson.
To proceed,  we need  to define the following matrix elements in
terms of decay constants and form factors:
\begin{eqnarray}\label{lep}
{\langle}0\mid  J _{\alpha\beta}^{D_2^*}(0) \mid
D_2^*(p',\epsilon)\rangle&=&m_{D_2^*}^3
f_{D_2^*}\epsilon_{\alpha\beta}\nonumber \\
\langle B(p)\mid J_{B}^{\dag}(0)\mid
 0\rangle&=&-i\frac{f_B m_B^2}{m_u+m_b}
\nonumber \\
{\langle}D_2^*(p',\epsilon)\mid J _{\mu}^{tr}(0) \mid
B(p)\rangle&=&h(q^2)\varepsilon_{\mu\nu\lambda\eta}\epsilon^{*^{\nu\theta}}P_{\theta}
P^{\lambda}q_{\eta}-iK(q^2)\epsilon^{*}_{\mu\nu}P^{\nu}\nonumber \\
&-&i\epsilon^{*}_{\lambda\eta}P^{\lambda}P^{\eta}\left[P_{\mu}b_{+}(q^2)+q_{\mu}b_{-}(q^2)
\right],
\end{eqnarray}
where $h(q^2)$, $K(q^2)$, $b_{+}(q^2)$ and $b_{-}(q^2)$ are transition form factors; and  $f_{D_2^*}$ and $f_B$ are leptonic decay constants of
$D_2^*$ and $B$ mesons, respectively. By combining Eqs. (\ref{phen1})
and  (\ref{lep}) and performing summation over polarization
tensors using
\begin{eqnarray}\label{polarizationt1}
\epsilon_{\alpha\beta}\epsilon_{\nu\theta}^*=\frac{1}{2}T_{\alpha\nu}T_{\beta\theta}+
\frac{1}{2}T_{\alpha\theta}T_{\beta\nu}
-\frac{1}{3}T_{\alpha\beta}T_{\nu\theta},
\end{eqnarray}
with
\begin{eqnarray}\label{polarizationt2}
T_{\alpha\nu}=-g_{\alpha\nu}+\frac{p'_\alpha
p'_\nu}{m_{D_2^*(2460)}^2},
\end{eqnarray}
the final representation of the physical side is obtained as
\begin{eqnarray}\label{phen2}
\Pi _{\mu\alpha\beta}^{phen}&=&\frac{f_{D_2^*}f_B m_{D_2^*} m_B^2}
{8(m_b+m_u)(p^2-m_B^2)(p'^2-m_{D_2^*}^2)}
\Bigg\{\frac{2}{3}\Big[-\Delta K(q^2)+\Delta'b_{-}(q^2)
\Big]q_{\mu}g_{\beta\alpha}\nonumber \\
 &+&\frac{2}{3}\Big[(\Delta-4m_{D_2^*}^2) K(q^2)+\Delta'b_{+}(q^2)
\Big]P_{\mu}g_{\beta\alpha}+i(\Delta-4m_{D_2^*}^2)
h(q^2)\varepsilon_{\lambda\nu\beta\mu}P_{\lambda}P_{\alpha}q_{\nu}
\nonumber \\
&+&\Delta K(q^2)q_{\alpha}g_{\beta\mu}+ \mbox{other
structures}\Bigg\}+...,
\end{eqnarray}
where
\begin{eqnarray}\label{phen2}
\Delta &=&m_B^2+3m_{D_2^*(2460)}^2-q^2,
\nonumber \\
\Delta'
&=&m_B^4-2m_B^2(m_{D_2^*(2460)}^2+q^2)+(m_{D_2^*(2460)}^2-q^2)^2.
\end{eqnarray}
We will use
the explicitly written  structures  to find the aforesaid form factors.

\subsection{The QCD side}

On the QCD  side, the correlation function is calculated by
expanding the time ordering product of the $B$ and $D_2^*(2460)$
mesons' currents and the transition current via operator product expansion (OPE) in
deep Euclidean region where the short (perturbative) and long distance (nonperturbative) contributions are
separated. 
By inserting the previously represented  currents  into Eq.
(\ref{correl.func.101})  and after contracting out all quark fields
applying the Wick's theorem, we obtain
\begin{eqnarray}\label{correl.func.2}
\Pi^{QCD} _{\mu\alpha\beta}(q^2)&=&\frac{-i^3}{4}\int d^{4}x\int
d^{4}ye^{-ip\cdot x}e^{ip'\cdot y}
\nonumber \\
&\times& \Bigg\{Tr\left[S_u^{ik}(x-y)\gamma_\alpha\olra{\cal
D}_{\beta}(y)
S_c^{ij}(y)\gamma_\mu(1-\gamma_5)S_b(-x)^{jk}\gamma_5\right]+
\left[\beta\leftrightarrow\alpha\right]
\Bigg\}.\nonumber\\
\end{eqnarray}
To proceed, we need the expressions of the heavy  and light quarks propagators.
Up to the terms considered in this study they are respectively given as
\begin{eqnarray}\label{heavypropagator}
S_{Q}^{ij}(x)=\frac{i}{(2\pi)^4}\int d^4k e^{-ik \cdot x} \left\{
\frac{\!\not\!{k}+m_c}{k^2-m_c^2}\delta_{ij}
 +\cdots\right\} \, ,
\end{eqnarray}
and
\begin{eqnarray}\label{lightpropagator}
S_{q}^{ij}(x)&=& i\frac{\!\not\!{x}}{ 2\pi^2 x^4}\delta_{ij}
-\frac{m_q}{4\pi^2x^2}\delta_{ij}-\frac{\langle
\bar{q}q\rangle}{12}\Big(1 -i\frac{m_q}{4}
\!\not\!{x}\Big)\delta_{ij} -\frac{x^2}{192}m_0^2\langle
\bar{q}q\rangle\Big(1-i\frac{m_q}{6} \!\not\!{x}\Big)\delta_{ij}
+\cdots \, .
\nonumber\\
\end{eqnarray}
After putting the expressions of the  quarks propagators and applying the
derivatives with respect to x and y in  Eq. (\ref{correl.func.2}),
the following  expression for the QCD side of the correlation
function in coordinate space is obtained:
\begin{eqnarray}\label{correl.func.3}
\Pi^{QCD}_{\mu\alpha\beta}(q^2)&=&\frac{i^5N_c}{4}
 \int\frac{d^4k}{(2\pi)^4}
 \int\frac{d^4k_1}{(2\pi)^4}\int
d^{4}xe^{-ip\cdot x}\int d^{4}ye^{ip'\cdot y} \frac{e^{-ik\cdot y}}{k^2-m_c^2}\frac{e^{ik_1\cdot x}}{k^2_1-m_b^2}\Bigg\{ik_{\beta}
\nonumber\\
&\times&Tr\Big[\Big(\frac{i(\!\not\!{x}-\!\not\!{y})}{2\pi^2(x-y)^4}-\frac{\langle\bar{u}u\rangle}{12}
-\frac{(x-y)^2}{192}m_0^2\langle
\bar{u}u\rangle\Big)\gamma_{\alpha}(\!\not\!{k}+m_c)\gamma_{\mu}(1-\gamma_5)(\!\not\!{k_1}+m_b)
\gamma_5\Big]
\nonumber\\
&+&Tr\Big[\Big(\frac{i}{2\pi^2}\Big(\frac{4(x-y)_{\beta}(\!\not\!{x}-\!\not\!{y})}{(x-y)^6}
-\frac{\gamma_{\beta}}{(x-y)^4}\Big)+\frac{(x-y)_{\beta}}{96}
m_0^2\langle
\bar{u}u\rangle\Big)\gamma_{\alpha}(\!\not\!{k}+m_c)\gamma_{\mu}
\nonumber\\
&\times& (1-\gamma_5)(\!\not\!{k_1}+m_b)
\gamma_5\Big]+\left[\beta\leftrightarrow\alpha\right] \Bigg\},
\end{eqnarray}
where $N_c=3$ is the color factor. In order to perform the
integrals, first the terms containing $\frac{1}{((x-y)^2)^n}$ are
transformed to the momentum space ($(x-y)\rightarrow t$), then 
the replacements $x_{\mu}\rightarrow i\frac{\partial}{\partial p_{\mu}}$ and 
 $y_{\mu}\rightarrow -i\frac{\partial}{\partial p'_{\mu}}$ are made. The four-integrals over $x$ and $y$
 give us two Dirac Delta functions which help us perform
the four-integrals over $k$ and $k_1$. The last four-integral over $t$
 is performed  using the Feynman
parametrization and 
\begin{eqnarray}\label{Int}
\int d^4t\frac{(t^2)^{\beta}}{(t^2+L)^{\alpha}}=\frac{i \pi^2
(-1)^{\beta-\alpha}\Gamma(\beta+2)\Gamma(\alpha-\beta-2)}{\Gamma(2)
\Gamma(\alpha)[-L]^{\alpha-\beta-2}}.
\end{eqnarray}
As a result,   the QCD side of the correlation function   is obtained in terms of the corresponding structures as
\begin{eqnarray}\label{QCDside}
\Pi^{QCD}_{\mu\alpha\beta}
(q^2)&=&\Big(\Pi^{pert}_1(q^2)+\Pi^{nonpert}_1(q^2)\Big)q_{\alpha}g_{\beta\mu}+
\Big(\Pi^{pert}_2(q^2)+\Pi^{nonpert}_2(q^2)\Big)q_{\mu}g_{\beta\alpha}\nonumber \\
&+&
\Big(\Pi^{pert}_3(q^2)+\Pi^{nonpert}_3(q^2)\Big)P_{\mu}g_{\beta\alpha}+
\Big(\Pi^{pert}_4(q^2)+\Pi^{nonpert}_4(q^2)\Big)\varepsilon_{\lambda\nu\beta\mu}P_{\lambda}
P_{\alpha}q_{\nu}
\nonumber \\
&+&other\,\,\, structures,
\end{eqnarray}
where, the perturbative parts
$\Pi^{pert}_i(q^2)$ are given in terms of double dispersion
integrals as
\begin{eqnarray}\label{QCDside}
\Pi^{pert}_i(q^2)=\int^{}_{}ds\int^{}_{}ds'
\frac{\rho_i(s,s',q^2)}{(s-p^2)(s'-p'^2)}.
\end{eqnarray}
The spectral densities $\rho_i(s,s',q^2)$ are
given by the imaginary parts of the $\Pi^{pert}_{i}(q^2)$
functions, i.e.,
$\rho_i(s,s',q^2)=\frac{1}{\pi}Im[\Pi^{pert}_{i}(q^2)]$. After lengthy calculations the  spectral densities corresponding to the selected structures are obtained as
\begin{eqnarray}\label{rho}
\rho_1(s,s',q^2)&=&\int_{0}^{1}dx\int^{1-x}_{0} dy
\Bigg\{\frac{1}{64\pi^2(x+y-1)^3}\Big[m_b(x+y-1)^3(8x^2-8y^2+6x-6y-6)
\nonumber \\
&+& 3m_c\Big(8x^5+6x^4(4y-3)
-6x(y-1)^2(3+2y+4y^2)-2(2+3y+4y^2)
\nonumber \\
&\times&
(y-1)^3+2x^3(1-18y+8y^2)+x^2(22-5y-16y^3)\Big)\Big]\Bigg\},
\nonumber \\
\rho_2(s,s',q^2)&=&\int_{0}^{1}dx\int^{1-x}_{0} dy
\Bigg\{\frac{-1}{32\pi^2(x+y-1)^3}\Big[m_b(x+y-1)^3(2x^2-2y^2+6x-6y-3)
\nonumber \\
&+&3m_c\Big(2x^5
-3x(y-1)^2(1+2y^2)-(y-1)^3(1+2y^2)+x^3(5-12y+4y^2)
\nonumber \\
&+&6x^4(y-1)+x^2(1+4y-4y^3)\Big)\Big]\Bigg\},
\nonumber \\
\rho_3(s,s',q^2)&=&\int_{0}^{1}dx\int^{1-x}_{0} dy
\Bigg\{\frac{1}{32\pi^2(x+y-1)^3}\Big[m_b(2x^2+2y^2+x(6+4y)+6y-3)
\nonumber \\
&\times&(x+y-1)^3+3m_c\Big(2x^5+2x^4(5y-3)+(y-1)^3(1+2y^2)+x(y-1)^2
\nonumber \\
&\times& (3-4y+10y^2)+x^3(7-24y+20y^2)
+x^2(20y^3-36y^2+20y-5)\Big)\Big]\Bigg\},
\nonumber \\
\rho_4(s,s',q^2)&=&0.
\end{eqnarray}

For the nonperturbative parts we get
\begin{eqnarray}\label{rho}
\Pi^{nonpert}_1(q^2)&=&\Bigg\{\frac{m_b^4+4m_b^2
m_c^2+2m_b^2(m_c^2-q^2)+(m_c^2-q^2)^2}{64r^2r'^2}+\frac{m_b^2
m_c^2(m_b^2+m_c^2-q^2)}{32 r^2 r'^3}
\nonumber \\
&+&\frac{m_b^3
m_c+m_b^2m_c^2+2m_bm_c^3+m_c^4-m_c^2q^2}{32rr'^3}-\frac{m_b^2+4m_b
m_c+m_c^2-q^2}{64rr'^2}
\nonumber \\
&+&\frac{m_b^4+2m_b^3m_c+m_b^2m_c^2-m_b^2q^2}{32r^3r'}+\frac{3m_b^2+2m_bm_c
+3m_c^2-3q^2}{64r^2r'}+\frac{m_b^2}{32r^3}
\nonumber \\
&+&\frac{m_c^2}{32r'^3}-
\frac{1}{32r'^2}+\frac{1}{32r^2}-\frac{1}{32rr'}
\Bigg\}m_0^2\langle
\bar{u}u\rangle
\nonumber \\
&-&
\Big(\frac{m_b^2+2m_bm_c+m_c^2-q^2}{16rr'}+\frac{1}{16r}+\frac{1}{16r'}\Big)\langle
\bar{u}u\rangle,
\nonumber \\
\Pi^{nonpert}_2(q^2)&=&0,
\nonumber \\
\Pi^{nonpert}_3(q^2)&=&\frac{m_0^2\langle \bar{u}u\rangle}{8rr'},
\nonumber \\
\Pi^{nonpert}_4(q^2)&=&-i\Big\{\frac{m_c^2}{32rr'^3}+\frac{m_b^2}{32r^3r'}+
\frac{m_b^2+m_c^2-q^2}{64r^2r'^2}-\frac{1}{32r^2r'}\Big\}m_0^2\langle
\bar{u}u\rangle+i\frac{\langle \bar{u}u\rangle}{16rr'}.
\end{eqnarray}
where $r=p^2-m_b^2$ and $r'=p'^2-m_c^2$.

To obtain sum rules for the form factors, the coefficients of the
same structures from both sides of the correlation functions are
matched. In order to suppress the contributions of the higher
states and continuum, we apply double Borel transformation with
respect to the initial and final  momenta squared using
\begin{eqnarray}\label{doubleBorel}
\widehat{B}\frac{1}{(p^2-m_b^2)^m}\frac{1}{(p'^2-m_c^2)^n}\rightarrow
\frac{(-1)^{m+n}}{\Gamma[m]
\Gamma[n]}e^{-m_b^2/M^2}e^{-m_c^2/M'^2}\frac{1}{(M^2)^{m-1}(M'^2)^{n-1}},
\end{eqnarray}
where $M^2$ and $M'^2$ are Borel mass parameters.
We also use the quark-hadron duality assumption, i.e.,
\begin{eqnarray}\label{duality}
\rho^{higher\,\,\,states}(s,s',q^2)=\rho^{OPE}(s,s',q^2)\theta(s-s_0)\theta(s'-s'_0), 
\end{eqnarray}
where $s_0$ and $s'_0$ are continuum thresholds in the initial and final mesonic channels, respectively.
After these procedures, the following sum rules for the form factors are
obtained:
\begin{eqnarray}\label{K}
K(q^2)&=&\frac{8(m_b+m_u)}{f_B
f_{D_2^*}m_{D_2^*}(m_B^2q^2-m_B^4-3m_B^2m_{D^*_2}^2)}e^{\frac{m_B^2}{M^2}}
e^{\frac{m_{D_2^*}^2}{M'^2}}
\nonumber \\
&&\Bigg\{\int^{s_0}_{(m_b+m_u)^2}ds
\int^{s_{0}^{'}}_{(m_c+m_u)^2}ds'\int_{0}^{1}dx \int_{0}^{1-x}dy
e^{\frac{-s}{M^2}}e^{\frac{-s'}{M'^2}}\Big[\frac{1}{256\pi^4(x+y-1)^3}
\nonumber \\
&& \Big(2m_b(x+y-1)^3(4x^2-4y^2+3x-3y-3)+3m_c\Big(8x^5+6x^4(4y-3)
\nonumber \\
&-&6x(y-1)^2(3+2y+4y^2)
-2(y-1)^3(2+3y+4y^2)+2x^3(1-18y+8y^2)
\nonumber \\
&+&x^2(22-5y-16y^3)\Big)\Big)\Big]\theta[L(s,s',q^2)]+e^{\frac{-m_b^2}{M^2}}e^{\frac{-m_c^2}{M'^2}}
\Big[\frac{\langle \bar{u}u\rangle}{16}\Big(m_b^2+2m_b
m_c+m_c^2-q^2\Big) \nonumber \\
&+&\frac{m_0^2\langle \bar{u}u\rangle}{64
}\Big(2+\frac{3m_b^2+2m_bm_c+3m_c^2-3q^2}{M^2}-\frac{m_b^2+4m_bm_c+m_c^2-q^2}{M'^2}
\nonumber \\
&-&\frac{m_b^4+2m_b^3m_c+m_b^2m_c^2-m_b^2q^2}{M^4}
-\frac{m_b^3m_c+m_b^2m_c^2+2m_bm_c^3+m_c^4-m_c^2q^2}{M'^4}
\nonumber \\
&-&\frac{m_b^4+4m_bm_c^3+2m_b^2m_c^2+m_c^4-m_c^2q^2-m_b^2q^2+q^4}{M^2M'^2}
+\frac{m_b^5m_c+m_bm_c^5-m_b^2m_c^2q^2}{M^2M'^4}
 \Big)\Big] \Bigg\},
\nonumber \\
b_{-}(q^2)&=&-\frac{12(m_b+m_u)}{f_B
f_{D_2^*}m_B^2m_{D_2^*}\Big(m_B^4+(m_{D_2^*}^2-q^2)^2-2m_B^2(m_{D_2^*}^2+q^2)\Big)}
e^{\frac{m_B^2}{M^2}} e^{\frac{m_{D_2^*}^2}{M'^2}}
\nonumber \\
&\times&\Bigg\{\int^{s_0}_{(m_b+m_u)^2}ds
\int^{s_{0}^{'}}_{(m_c+m_u)^2}ds'\int_{0}^{1}dx \int_{0}^{1-x}dy
e^{\frac{-s}{M^2}}e^{\frac{-s'}{M'^2}}\Big[\frac{1}{128\pi^4(x+y-1)^3}
\nonumber \\
&\times&\Big(m_b(x+y-1)^3(3-6x-2x^2+6y+2y^2)-3m_c\Big(6x^4(y-1)-3x(y-1)^2(1+2y^2)
\nonumber \\
&-&
(y-1)^3(1+2y^2)+x^3(5-12y+4y^2)+x^2(1+4y-4y^3)+2x^5\Big)\Big)\Big]\theta[L(s,s',q^2)]
\nonumber \\
&-&e^{\frac{-m_B^2}{M^2}} e^{\frac{-m_{D_2^*}^2}{M'^2}}\frac{f_B
f_{D_2^*}m_B^2m_{D_2^*}(m_B^2+3m_{D_2^*}^2+q^2)}{12(m_b+m_u)}K(q^2)
\Bigg\},
\nonumber \\
b_{+}(q^2)&=&-\frac{12(m_b+m_u)}{f_B
f_{D_2^*}m_B^2m_{D_2^*}\Big(m_B^4+(m_{D_2^*}^2-q^2)^2-2m_B^2(m_{D_2^*}^2+q^2)\Big)}
e^{\frac{m_B^2}{M^2}} e^{\frac{m_{D_2^*}^2}{M'^2}}
\nonumber \\
&\times&\Bigg\{\int^{s_0}_{(m_b+m_u)^2}ds
\int^{s_{0}^{'}}_{(m_c+m_u)^2}ds'\int_{0}^{1}dx \int_{0}^{1-x}dy
e^{\frac{-s}{M^2}}e^{\frac{-s'}{M'^2}}\Big[\frac{1}{128\pi^4(x+y-1)^3}
\nonumber \\
&\times&\Big(m_b(x+y-1)^3(2x^2+2y^2+6x+6y+4xy-3)+3m_c\Big(2x^5-6x^4+10x^4y
\nonumber \\
&+&(y-1)^3
(1+2y^2)+x(y-1)^2(3-4y+10y^2)+x^2(20y^3-36y^2+20y-5)
\nonumber \\
&+&x^3(7-24y+20y^2)\Big)\Big)\Big]\theta[L(s,s',q^2)]-\frac{m_0^2\langle
\bar{u}u\rangle}{8}e^{\frac{-m_b^2}{M^2}}e^{\frac{-m_c^2}{M'^2}}
\nonumber \\
&-&e^{\frac{-m_B^2}{M^2}} e^{\frac{-m_{D_2^*}^2}{M'^2}}\frac{f_B
f_{D_2^*}m_B^2m_{D_2^*}(m_{D_2^*}^2-m_B^2+q^2)}{12(m_b+m_u)}K(q^2)
\Bigg\},
\nonumber \\
h(q^2)&=&\frac{8(m_b+m_u)}{f_B
f_{D_2^*}m_B^2m_{D_2^*}\Big(m_{D_2^*}^2-m_B^2+q^2\Big)}e^{\frac{m_B^2}{M^2}}
e^{\frac{m_{D_2^*}^2}{M'^2}}e^{\frac{-m_b^2}{M^2}}e^{\frac{-m_c^2}{M'^2}}\Big\{
-\frac{\langle\bar{u}u\rangle}{16}
\nonumber \\
&+& \frac{m_0^2\langle
\bar{u}u\rangle}{64}\Big[\frac{2}{M^2}+\frac{2}{M'^2}+\frac{m_b^2}{M^4}+\frac{m_c^2}{M'^4}
+\frac{m_b^2-m_c^2+q^2}{M^2M'^2}\Big]\Big\},
\end{eqnarray}
where
\begin{eqnarray}\label{L}
L(s,s',q^2)=s'x-s'x^2-m_c^2x-m_b^2y+sy+q^2xy-sxy-s'xy-sy^2.
\end{eqnarray}

\section{Numerical results and discussions}
\begin{table}[ht]\label{table1}
\centering \rowcolors{1}{lightgray}{white}
\begin{tabular}{cc}
\hline \hline
   Parameters  &  Values
           \\
\hline \hline
$m_{c}$              & $(1.275\pm0.025)~ GeV$\\
$m_{b}$              & $(4.65\pm0.03)~ GeV$\\
$ m_e $              &   $ 0.00051  $ $GeV$ \\
$ m_\mu $            &   $ 0.1056   $ $GeV$ \\
$ m_\tau $           &   $ 1.776 $  $GeV$ \\
$ m_{D_2^*(2460)}$    &   $ (2.4626\pm0.0007) $ $GeV$   \\
$ m_{B} $      &   $ (5.27925\pm0.00017) $ $GeV$   \\
$ f_{B} $      &   $(210\pm 40) ~MeV $   \\
$ f_{D_2^*(2460)} $      &   $0.0317\pm0.0092 $   \\
$ G_{F} $            &   $ 1.17\times 10^{-5} $ $GeV^{-2}$ \\
$  V_{cb} $ &   $ (41.2\pm1.1)\times 10^{-3} $   \\
$ \langle0|\overline{u}u(1GeV)|0\rangle$        &   $ -(0.24\pm0.01)^3$ $GeV^{3}$   \\
$ m_0^2(1GeV) $       &   $(0.8\pm0.2)$ $GeV^{2}$   \\
$ \tau_B $       &   $(1641\pm8)\times10^{-15}s$   \\
 \hline \hline
\end{tabular}
\caption{Input parameters used in  calculations
\cite{Beringer,Sundu,Na,Ioffe,Dosch}.}
\end{table}
In this part, we numerically analyze the obtained sum rules for the form factors in the previous section and obtain  their variations in terms of $q^2$. 
For this aim we need some  input parameters whose values  are given in Table 1. Besides these input parameters,
the sum rules for the form factors contain four auxiliary
parameters, namely the Borel mass parameters $M^2$ and $M'^2$ and
continuum thresholds $s_0$ and $s'_0$. We shall find their working regions such that the form factors weakly depend on these parameters. The continuum thresholds
are not completely arbitrary but they are related to the energy of
the first excited state in initial and final mesonic channels. Our calculations show that in the intervals $31~GeV^2\leq
s_0\leq35~GeV^2$ and $7~GeV^2\leq
s'_0\leq9~GeV^2$, our results weakly depend on the continuum thresholds. The
working regions for the Borel mass parameters are determined by
requiring that not only the contributions of the higher states and
continuum are sufficiently suppressed but also the contributions
of the operators with higher dimensions are relatively small, i.e., the series of sum rules for the form factors are convergent. As a result, we
find the working regions $10~GeV^2\leq M^2\leq 20GeV^2$ and
$5GeV^2\leq M'^2\leq 15 GeV^2$. To show how  the form factors depend on the auxiliary parameters, as examples,  we depict the variations of the form factors $K(q^2)$ and  $b_+(q^2)$ at $q^2=0$ with respect to the variations 
of the related auxiliary parameters in their working regions in figures 1 and 2. From these figures, we see that the form factors weakly depend on the auxiliary parameters in their working regions.
\begin{figure}[h!]
\begin{center}
\includegraphics[totalheight=7cm,width=7cm]{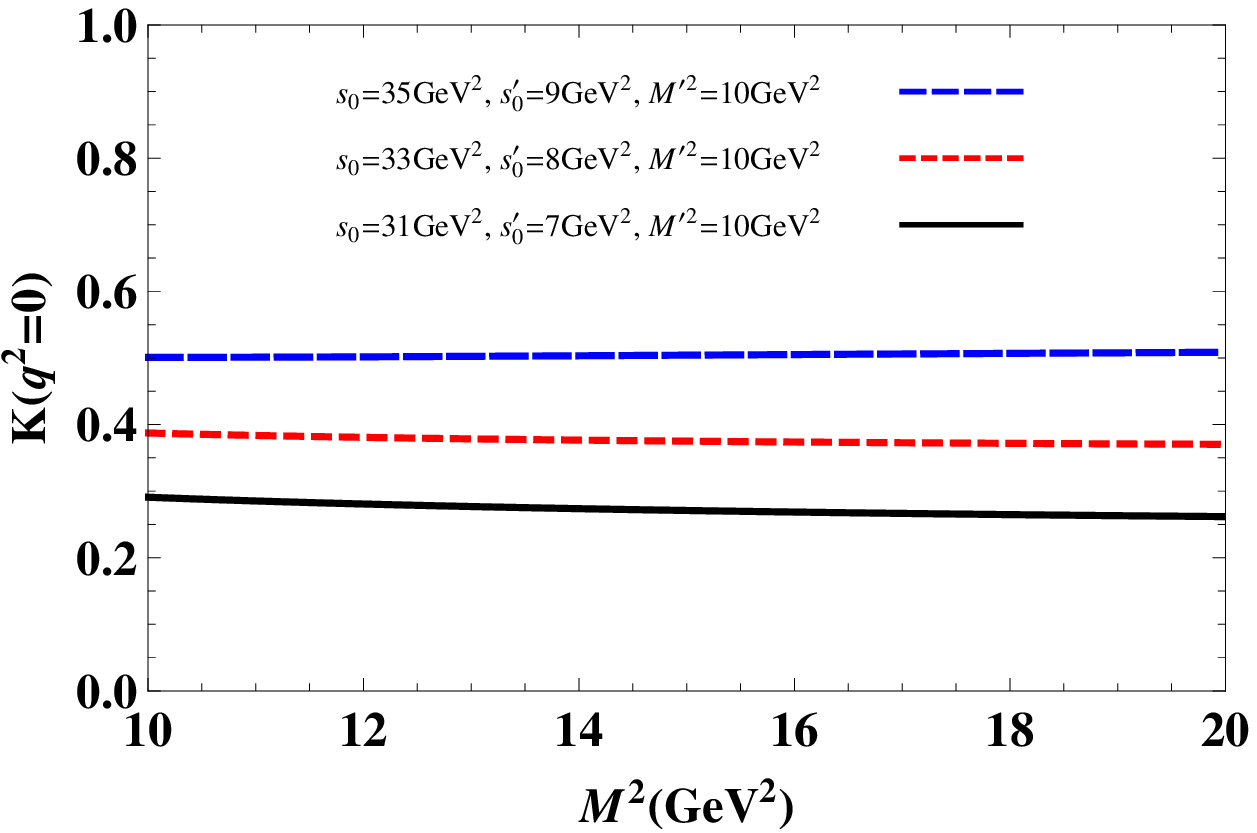}
\includegraphics[totalheight=7cm,width=7cm]{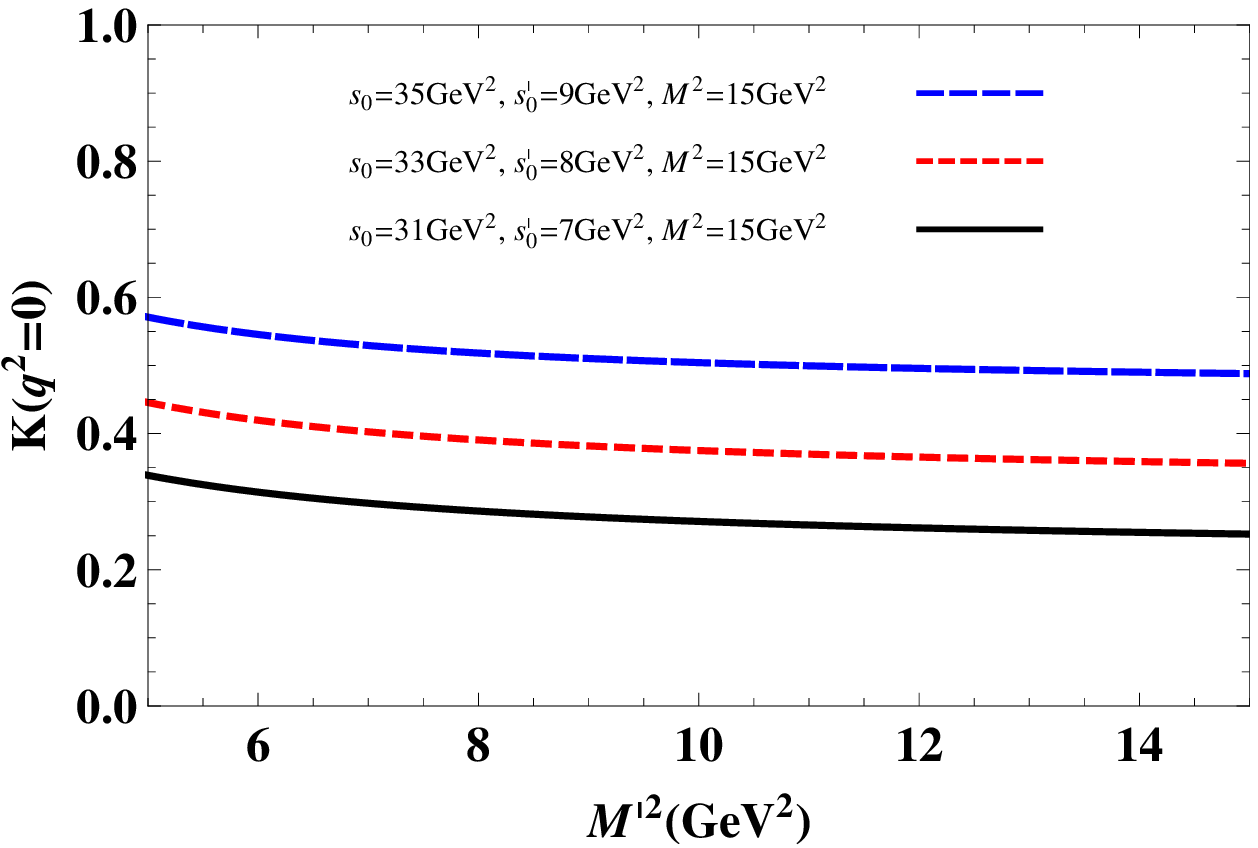}
\end{center}
\caption{\textbf{Left:} K($q^2=0$) as a function of the Borel mass
$M^2$ at fixed values of the $s_0$, $s'_0$ and $M^{\prime^2}$. \textbf{Right:}
 K($q^2=0$) as a function of the
Borel mass $M^{\prime^2}$ at fixed values of the $s_0$, $s'_0$ and $M^2$. } \label{KMsqMpsq}
\end{figure}
%
%
\begin{figure}[h!]
\begin{center}
\includegraphics[totalheight=7cm,width=7cm]{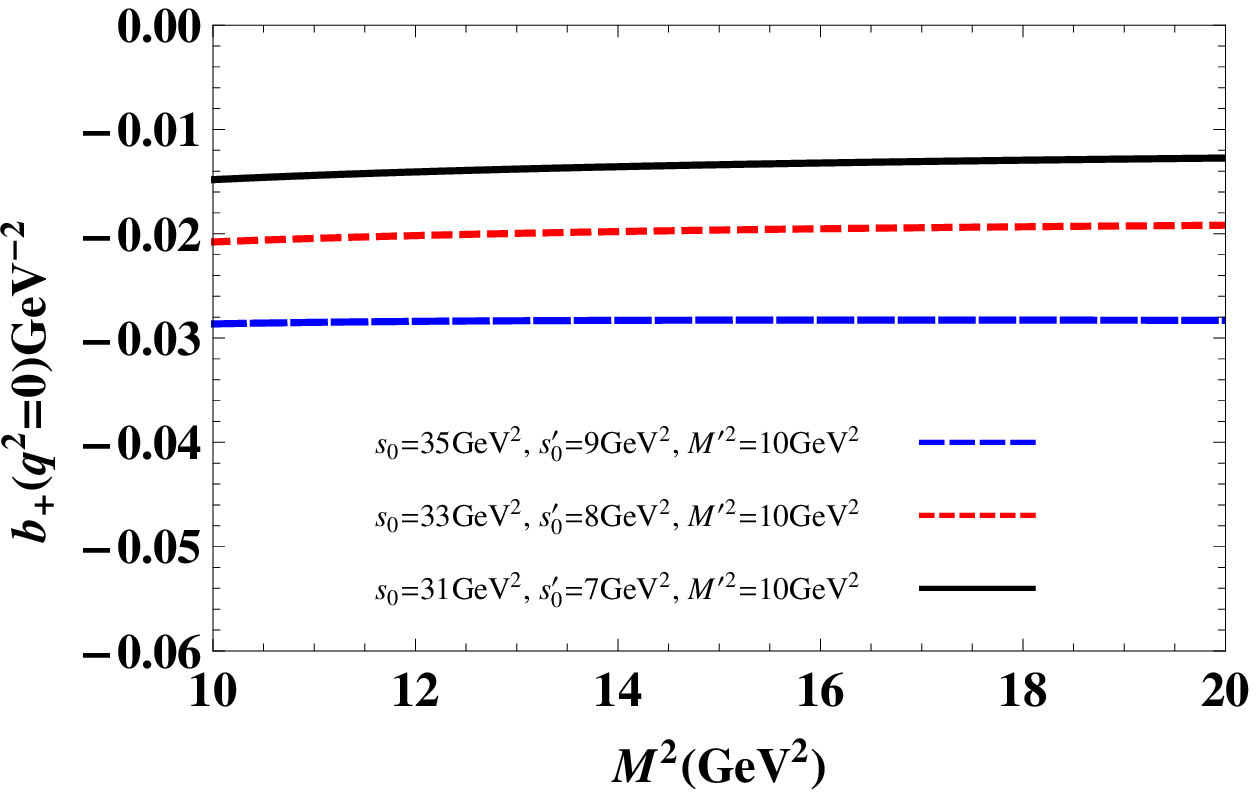}
\includegraphics[totalheight=7cm,width=7cm]{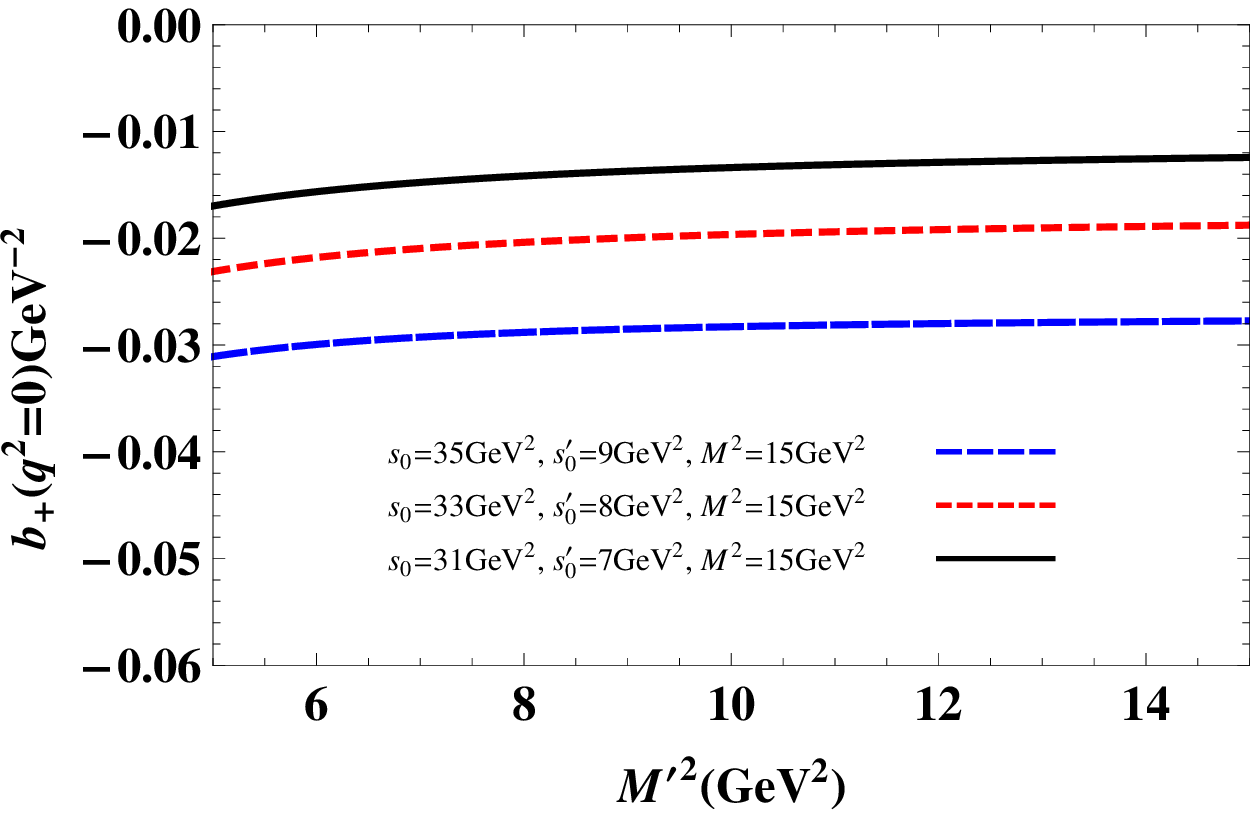}
\end{center}
\caption{\textbf{Left:} $b_{+}$($q^2=0$) as a function of the
Borel mass $M^2$ at fixed values of the $s_0$, $s'_0$ and $M^{\prime^2}$. \textbf{Right:}
 $b_{+}$($q^2=0$) as a function of the
Borel mass $M^{\prime^2}$ at fixed values of the $s_0$, $s'_0$ and $M^2$. } \label{bplusMsqMpsq}
\end{figure}
%
%
%
\begin{table}[h]
\renewcommand{\arraystretch}{1.5}
\addtolength{\arraycolsep}{3pt}
$$
\begin{array}{|c|c|c|c|c|}
\hline \hline
         &f_0 & c_1 & c_2 & m_{fit}^2   \\
\hline
  \mbox{$K (q^2)$} &0.54\pm0.14&0.70\pm0.07&0.41\pm0.02&27.88\pm0.01 \\
  \hline
  \mbox{$b_{-} (q^2)$} &0.007\pm0.002~GeV^{-2}&0.14\pm0.04&10.70\pm0.82&27.88\pm0.01 \\
  \hline
  \mbox{$b_{+} (q^2)$} &-0.03\pm0.01~GeV^{-2}&1.20\pm0.15&22.52\pm1.68&27.88\pm0.01 \\
  \hline
  \mbox{$h (q^2)$} &-0.010\pm0.003~GeV^{-2}&1.19\pm0.13&1.12\pm0.08&27.88\pm0.01 \\
                    \hline \hline
\end{array}
$$
\caption{Parameters appearing in the fit function 1 of the form
factors.} \label{fitfunction1}
\renewcommand{\arraystretch}{1}
\addtolength{\arraycolsep}{-1.0pt}
\end{table}

\begin{figure}[h!]
\begin{center}
\includegraphics[totalheight=7cm,width=7cm]{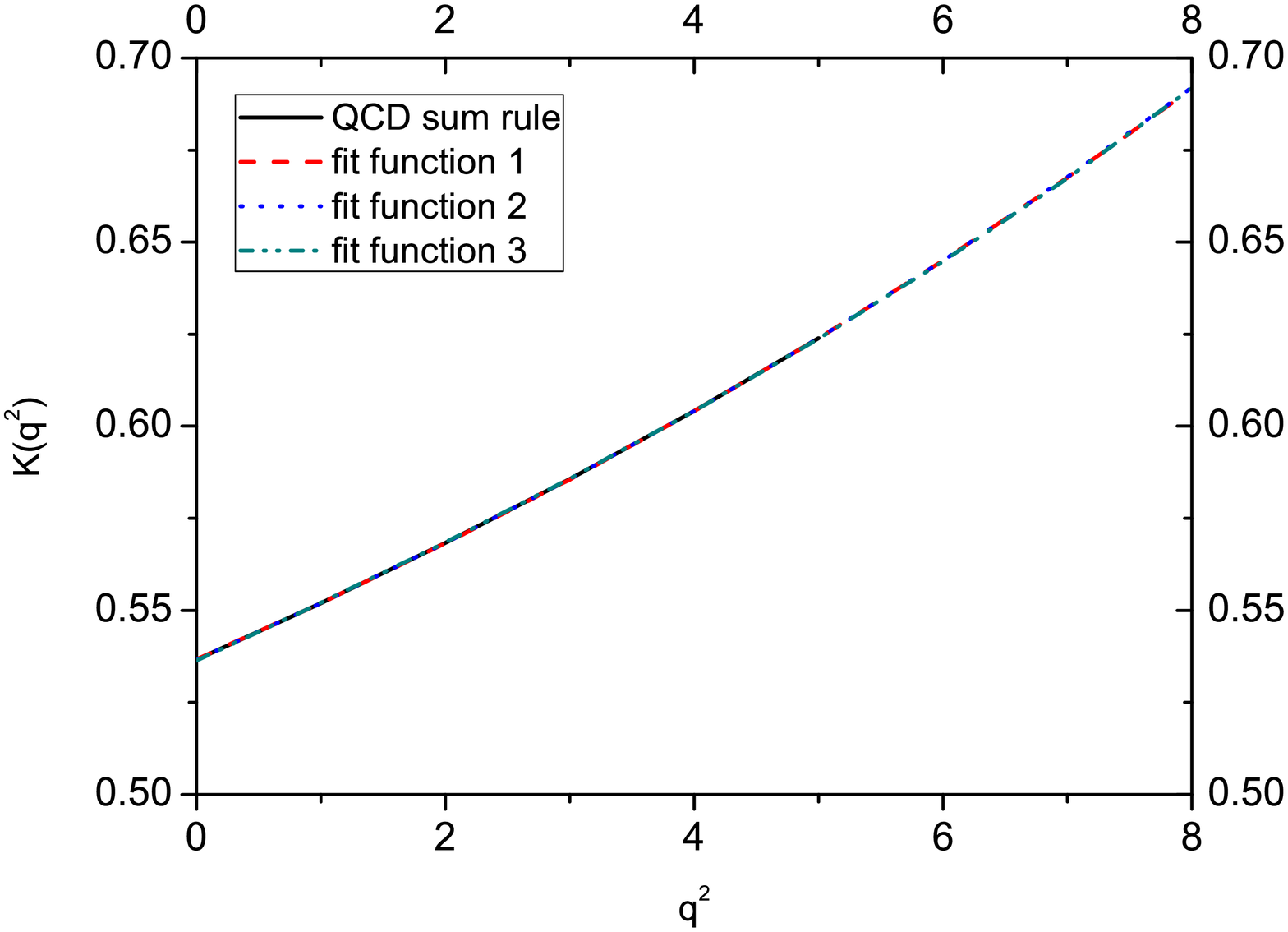}
\includegraphics[totalheight=7cm,width=7cm]{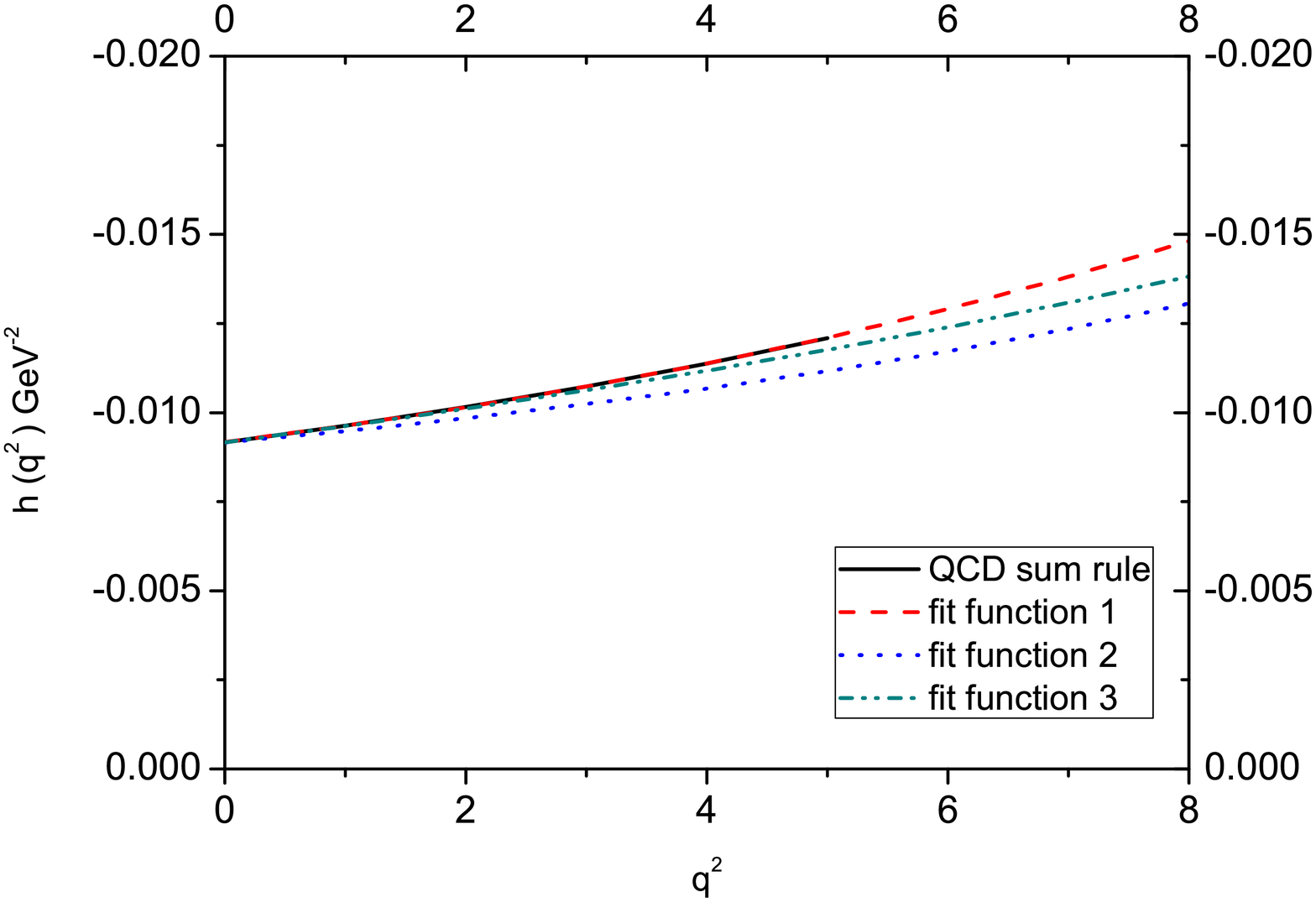}
\end{center}
\caption{\textbf{Left:} K($q^2$) as a function of $q^2$ at
$M^2=15GeV^2$, $M^{\prime^2}=10GeV^2$, $s_0=35GeV^2$ and
$s_{0}^{\prime}=9GeV^2$ . \textbf{Right:}
 h($q^2$) as a function of $q^2$ at
$M^2=15GeV^2$, $M^{\prime^2}=10GeV^2$, $s_0=35GeV^2$ and
$s_{0}^{\prime}=9GeV^2$. } \label{Kandhqsq}
\end{figure}
\begin{figure}[h!]
\begin{center}
\includegraphics[totalheight=7cm,width=7cm]{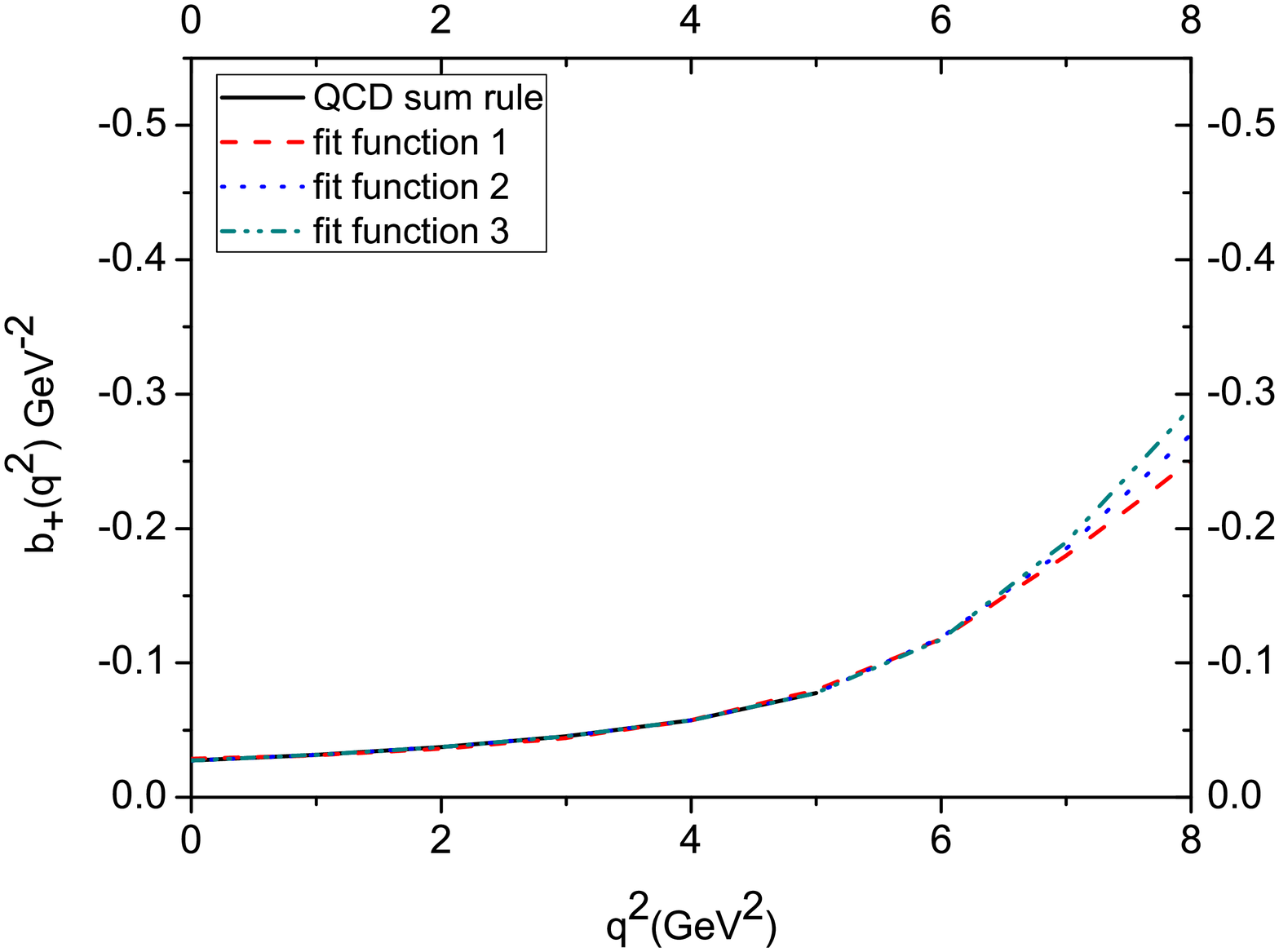}
\includegraphics[totalheight=7cm,width=7cm]{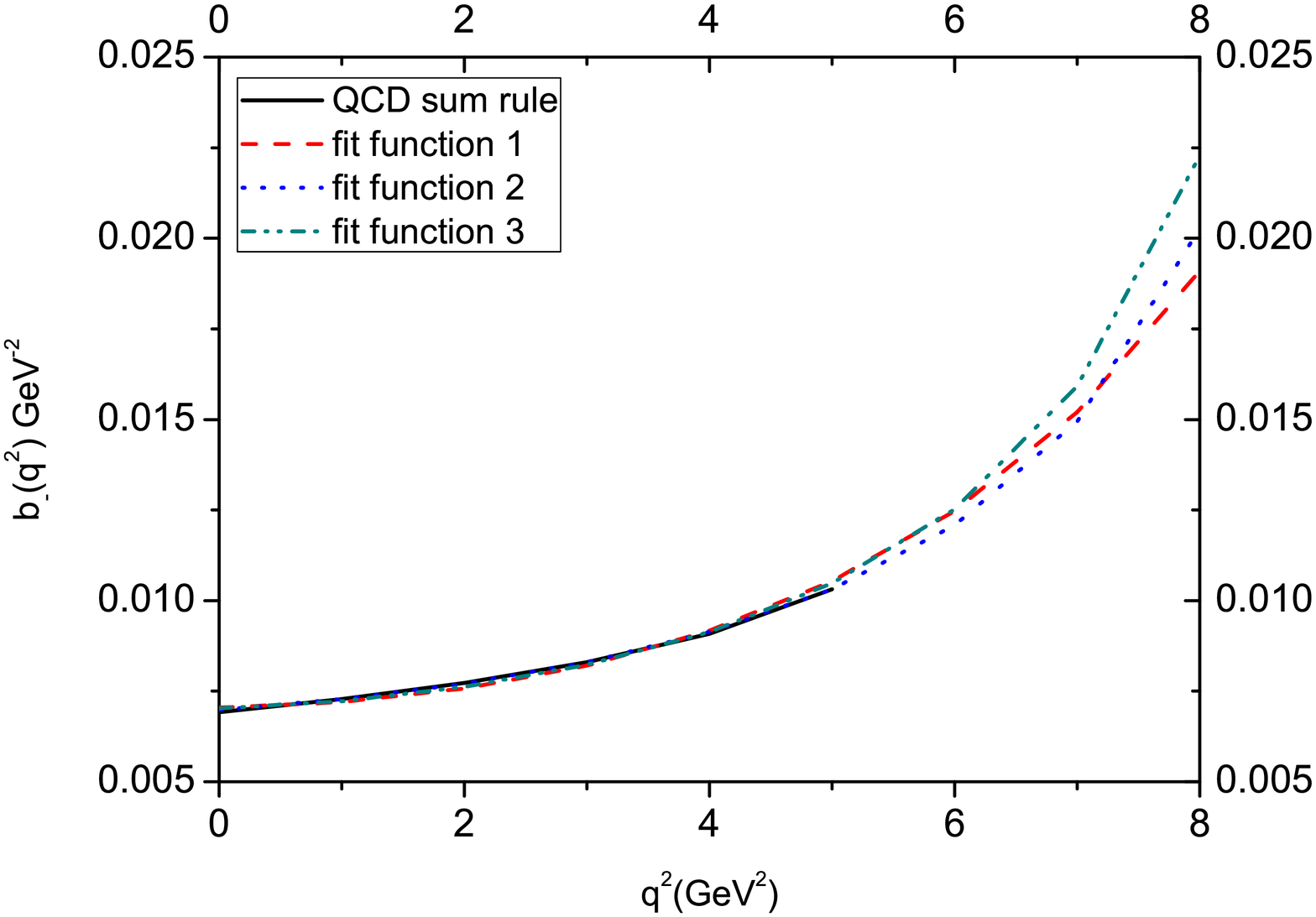}
\end{center}
\caption{\textbf{Left:} $b_{+}$($q^2$) as a function of $q^2$ at
$M^2=15GeV^2$, $M^{\prime^2}=10GeV^2$, $s_0=35GeV^2$ and
$s_{0}^{\prime}=9GeV^2$ . \textbf{Right:}
 $b_{-}$($q^2$) as a function of $q^2$ at
$M^2=15GeV^2$, $M^{\prime^2}=10GeV^2$, $s_0=35GeV^2$ and
$s_{0}^{\prime}=9GeV^2$. } \label{bminusandbplusqsq}
\end{figure}
%
%
%
%

Using the  working regions for the continuum thresholds and
Borel mass parameters as well as other input parameters we proceed to find the behavior of the form factors in terms of $q^2$. Our
calculations show that the form factors are truncated at
$q^2\simeq5GeV^2$. In order to estimate the decay width of the $B
\rightarrow D_2^*(2460)\ell\overline{\nu}$ transition, we have to
obtain their fit functions in the whole physical region,
$m_{\ell}^2\leq q^2\leq(m_B-m_{D_2^*})^2$. We find that the sum
rules predictions for the form factors are well fitted to the
following function:
\begin{eqnarray}\label{fitfunc}
f(q^2)=f_0
\exp \Big[c_1\frac{q^2}{m_{fit}^2}+c_2\Big(\frac{q^2}{m_{fit}^2}\Big)^2
\Big]
\end{eqnarray}
where, the values of the parameters $f_0$, $c_1$, $c_2$ and
$m_{fit}^2$  are
presented in Table \ref{fitfunction1}. In the following, we will recall this parametrization as \textit{fit function 1}. To compare our results with other parametrization, we also use the following fit functions to extrapolate 
the form factors to whole physical regions (see \cite{Wei,Wei2,Wei3,Wei4}):
\begin{itemize}
 \item \textit{fit function 2}
\begin{eqnarray}\label{fitfunc2}
f(q^2)=\frac{f_0}{1-a(\frac{q^2}{m_{B}^2})+b(\frac{q^2}{m_B^2})^2},
\end{eqnarray}
\item \textit{fit function 3}
\begin{eqnarray}\label{fitfunc3}
f(q^2)=\frac{f_0}{\Big(1-\frac{q^2}{m_{B}^2}\Big)\Big[1-A(\frac{q^2}{m_{B}^2})+B(\frac{q^2}{m_B^2})^2\Big]},
\end{eqnarray}
\end{itemize}
where the parameters $a$, $b$, $A$ and $B$ and the values of corresponding form factors at $q^2=0$ are given in Tables \ref{fitfunction2} and \ref{fitfunction3}, respectively.
\begin{table}[h]
\renewcommand{\arraystretch}{1.5}
\addtolength{\arraycolsep}{3pt}
$$
\begin{array}{|c|c|c|c|}
\hline \hline
         &f_0 & a & b    \\
\hline
  \mbox{$K (q^2)$} &0.54\pm0.14&0.75\pm0.03&-0.014\pm0.006 \\
  \hline
  \mbox{$b_{-} (q^2)$} &0.007\pm0.002~GeV^{-2}&0.95\pm0.04&-3.14\pm1.34 \\
  \hline
  \mbox{$b_{+} (q^2)$} &-0.03\pm0.01~GeV^{-2}&1.41\pm0.06&-4.63\pm2.05 \\
  \hline
  \mbox{$h (q^2)$} &-0.010\pm0.003~GeV^{-2}&1.27\pm0.05&0.058\pm0.002 \\
                    \hline \hline
\end{array}
$$
\caption{Parameters appearing in the fit function 2 of the form
factors.} \label{fitfunction2}
\renewcommand{\arraystretch}{1}
\addtolength{\arraycolsep}{-1.0pt}
\end{table}
\begin{table}[h]
\renewcommand{\arraystretch}{1.5}
\addtolength{\arraycolsep}{3pt}
$$
\begin{array}{|c|c|c|c|}
\hline \hline
         &f_0 & A & B    \\
\hline
  \mbox{$K (q^2)$} &0.54\pm0.14&-0.15\pm0.06&0.31\pm0.03 \\
  \hline
  \mbox{$b_{-} (q^2)$} &0.007\pm0.002~GeV^{-2}&-0.36\pm0.16&-7.72\pm0.86 \\
  \hline
  \mbox{$b_{+} (q^2)$} &-0.03\pm0.01~GeV^{-2}&1.89\pm0.81&-2.39\pm0.27 \\
  \hline
  \mbox{$h (q^2)$} &-0.010\pm0.003~GeV^{-2}&0.25\pm0.10&-0.35\pm0.04 \\
                    \hline \hline
\end{array}
$$
\caption{Parameters appearing in the fit function 3 of the form
factors.} \label{fitfunction3}
\renewcommand{\arraystretch}{1}
\addtolength{\arraycolsep}{-1.0pt}
\end{table}

The dependences of form factors on $q^2$ at different fixed values of auxiliary parameters are depicted in figures 3 and 4. These figures include the sum rules results (up to the truncated point) as well as the results
obtained using the above mentioned three different fit functions. From these figures it is clear that, in the case of the form factors $K(q^2)$, $b_+(q^2)$ and $b_-(q^2)$, all three fit functions reproduce the sum rules results up to
the truncated point, however, we see small differences between the predictions of these fit functions at higher values of $q^2$ except for the form factor $K(q^2)$ that all fit functions give the same results. 
In the case of the form factor $h(q^2)$, the parametrization 1 well fits to the sum rule result, 
but we see considerable differences of prediction of this parametrization  with those of fit functions 2 and 3, especially at higher values of $q^2$.

Now we proceed to  calculate the decay width and branching ratio
of the process under consideration. The differential decay width
for $B \rightarrow D_2^*(2460)\ell\overline{\nu}$ transition is
obtained as \cite{Wang}
\begin{eqnarray}\label{decaywidth}
\frac{d\Gamma}{dq^2}&=&\frac{\lambda(m_B^2,m_{D_2^*}^2,q^2)}{4m_{D_2^*}^2}
\Big(\frac{q^2-m_{\ell}^2}{q^2}\Big)^2\frac{\sqrt{\lambda(m_B^2,m_{D_2^*}^2,q^2)}G_F^2
V_{cb}^2}{384m_B^3\pi^3}\Bigg\{\frac{1}{2q^2}\Bigg[3m_{\ell}^2\lambda(m_B^2,m_{D_2^*}^2,q^2)
[V_0(q^2)]^2
\nonumber \\
&+&(m_{\ell}^2+2q^2)\Big|\frac{1}{2m_{D_2^*}}\Big[(m_B^2-m_{D_2^*}^2-q^2)(m_B-m_{D_2^*})V_1(q^2)
-\frac{\lambda(m_B^2,m_{D_2^*}^2,q^2)}{m_B-m_{D_2^*}}V_2(q^2)\Big]\Big|^2\Bigg]
\nonumber \\
&+&\frac{2}{3}(m_{\ell}^2+2q^2)\lambda(m_B^2,m_{D_2^*}^2,q^2)\Bigg[\Big|
\frac{A(q^2)}{m_B-m_{D_2^*}}-\frac{(m_B-m_{D_2^*})V_1(q^2)}{\sqrt{\lambda(m_B^2,m_{D_2^*}^2,q^2)}}
\Big|^2
\nonumber \\
&+&\Big|
\frac{A(q^2)}{m_B-m_{D_2^*}}+\frac{(m_B-m_{D_2^*})V_1(q^2)}{\sqrt{\lambda(m_B^2,m_{D_2^*}^2,q^2)}}
\Big|^2\Bigg]\Bigg\},
\end{eqnarray}
where
\begin{eqnarray}\label{decaywidth1}
A(q^2)&=&-(m_B-m_{D_2^*})h(q^2),
\nonumber \\
V_1(q^2)&=&-\frac{K(q^2)}{m_B-m_{D_2^*}},
\nonumber \\
V_2(q^2)&=&(m_B-m_{D_2^*})b_{+}(q^2),
\nonumber \\
V_0(q^2)&=&\frac{m_B-m_{D_2^*}}{2m_{D_2^*}}V_1(q^2)-\frac{m_B+m_{D_2^*}}{2m_{D_2^*}}V_2(q^2)
-\frac{q^2}{2m_{D_2^*}}b_{-}(q^2), \nonumber\\
\lambda(a,b,c)&=&a^2+b^2+c^2-2ab-2ac-2bc.
\end{eqnarray}

After performing integration over $q^2$ in Eq. (\ref{decaywidth})
in the interval $m_{\ell}^2\leq q^2 \leq (m_B-m_{D_2^*})^2$, we
obtain the total decay widths and branching ratios for all leptons and three different fit functions presented in Table
\ref{numresult}. The errors in the results belong to the uncertainties in  determination of the working regions for the auxiliary parameters as well as 
errors in the other input parameters.  From this Table, it is clear that, for the $e$ and  $\mu$ channels, all fit functions give roughly the same results. In the case of $\tau$, the fit functions 2 and 3
have approximately the same predictions, but they give results roughly \%38 smaller than that of the fit function 1. As it is expected, the values for the branching ratios in 
the cases of $e$ and $\mu$ are very close to each other for all fit functions. 
The orders of branching fractions  show that this transition  can be
detected at LHCb for all lepton channels. Note that  there are experimental data on 
 the products of branching fractions
for the decay chain  ${\cal{B}} (B\rightarrow
D_2^*\ell\overline{\nu}){\cal{B}} (D_2^*\rightarrow D\pi) $ provided by Belle \cite{belle} and BaBar \cite{babar1,babar2} Collaborations:
\begin{eqnarray}
 {\cal{B}} (B^+\rightarrow
\overline{D}_2^*\ell^{'+}\overline{\nu}_{\ell'}){\cal{B}} (\overline{D}_2^*\rightarrow D\pi)&=&2.2\pm0.3\pm0.4~~~~~~~~~~\mbox{Belle \cite{belle}},\nonumber\\
{\cal{B}} (B^+\rightarrow
\overline{D}_2^*\ell^{'+}\overline{\nu}_{\ell'}){\cal{B}} (\overline{D}_2^*\rightarrow D\pi)&=&1.4\pm0.2\pm0.2~~~~~~~~~~\mbox{BaBar \cite{babar1,babar2}}.
\end{eqnarray}
where $l'=e$ or $\mu$. Considering the recent experimental progress especially at LHC we hope we will have experimental data on the branching fraction
 of the semileptonic $B \rightarrow D_2^*(2460)\ell\overline{\nu}$ transition in near future, comparison of which with the results of the present work can give more information
about the nature and internal structure of the  $D_2^*(2460)$ tensor meson.
\begin{table}[h]
\renewcommand{\arraystretch}{1.5}
\addtolength{\arraycolsep}{3pt}
$$
\begin{array}{|c|c|c|}
\hline \hline
     \mbox{ fit function 1} &\Gamma(GeV) &   Br   \\
\hline
  \mbox{$B \rightarrow D_2^*(2460)\tau\overline{\nu}_{\tau}$} &(6.52\pm2.20)\times 10^{-17}&(0.16\pm0.06)\times 10^{-3} \\
  \hline
  \mbox{$B \rightarrow D_2^*(2460)\mu\overline{\nu}_{\mu}$} &(4.04\pm1.18)\times 10^{-16}&(1.00\pm0.29)\times 10^{-3} \\
  \hline
  \mbox{$B \rightarrow D_2^*(2460)e\overline{\nu}_{e}$} &(4.05\pm1.19)\times 10^{-16}&(1.01\pm0.30)\times 10^{-3} \\
                      \hline \hline
 \mbox{ fit function 2} &\Gamma(GeV) &   Br   \\
\hline
  \mbox{$B \rightarrow D_2^*(2460)\tau\overline{\nu}_{\tau}$} &(4.09\pm1.28)\times 10^{-17}&(0.10\pm0.03)\times 10^{-3} \\
  \hline
  \mbox{$B \rightarrow D_2^*(2460)\mu\overline{\nu}_{\mu}$} &(4.06\pm1.26)\times 10^{-16}&(1.01\pm0.32)\times 10^{-3} \\
  \hline
  \mbox{$B \rightarrow D_2^*(2460)e\overline{\nu}_{e}$} &(4.08\pm1.28)\times 10^{-16}&(1.02\pm0.32)\times 10^{-3} \\
                      \hline \hline
 \mbox{ fit function 3} &\Gamma(GeV) &   Br   \\
\hline
  \mbox{$B \rightarrow D_2^*(2460)\tau\overline{\nu}_{\tau}$} &(4.80\pm1.60)\times 10^{-17}&(0.12\pm0.04)\times 10^{-3} \\
  \hline
  \mbox{$B \rightarrow D_2^*(2460)\mu\overline{\nu}_{\mu}$} &(4.18\pm1.32)\times 10^{-16}&(1.04\pm0.34)\times 10^{-3} \\
  \hline
  \mbox{$B \rightarrow D_2^*(2460)e\overline{\nu}_{e}$} &(4.20\pm1.32)\times 10^{-16}&(1.05\pm0.34)\times 10^{-3} \\
                      \hline \hline
\end{array}
$$
\caption{Numerical results for the decay widths and branching ratios at
different lepton channels for different fit functions.} \label{numresult}
\renewcommand{\arraystretch}{1}
\addtolength{\arraycolsep}{-1.0pt}
\end{table}

At the end of this section we would like to calculate the ratio of the branching fraction in the case of  $\tau$ to that of the $e$ or $\mu$. From our calculations we obtain that
\begin{eqnarray}
{\cal R}=\frac{B \rightarrow D_2^*(2460)\tau\overline{\nu}_{\tau}}{B \rightarrow D_2^*(2460)\ell'\overline{\nu}_{\ell'}}
 = \left\{ \begin{array}{c}
0.16\pm0.04 ~~~~\mbox{fit function 1}, \\
0.10\pm0.02 ~~~~\mbox{fit function 2}, \\
 0.11\pm0.02 ~~~~\mbox{fit function 3}.
\end{array} \right.
\end{eqnarray}
 As we previously mentioned the  SM predictions in the $B$ to pseudoscalar and vector charmed mesons deviate at the  level of 3.4$~\sigma$ from the
experimental data. Our result on ${\cal R}$ in the case of tensor charmed current can be checked in future experiments. Comparison of the experimental data with the result of 
this work will illustrate whether these anomalous in the pseudoscalar and vector channels exist also in the tensor channel or not. 


\end{document}